\documentclass[letterpaper, 10 pt, conference]{ieeeconf} 

\usepackage{ifthen}
\def\TU{true}
\def\short{true}

\IEEEoverridecommandlockouts 
\usepackage{layout}

\usepackage{graphicx} 
\usepackage{epsfig} 
\usepackage{mathptmx} 
\usepackage{times} 
\usepackage{amsmath} 
\usepackage{amssymb} 
\usepackage{tikz}
\usepackage{pgfplots}
\usetikzlibrary{positioning,arrows}
\usetikzlibrary{shapes,arrows}
\usepackage{comment}
\newtheorem{thm}{Theorem}
\usepackage{xcolor}
\usepackage{fancyhdr}
\usepackage{cuted}

\title{\LARGE \bf

Event-Triggered $\ell_2$-Optimal Formation Control  with \\ State-Estimation  
for Agents Modeled as LPV Systems
}
\author{Gerald Gebhardt, Hamideh Saadabadi and Herbert Werner
\thanks{Herbert Werner is with the Institute of Control Systems, Hamburg University of Technology, Eissendorfer Str. 40, 21073 Hamburg, Germany
 {\tt\small h.werner@tuhh.de}}%
\thanks{Hamideh Saadabadi is with the Institute of Control Systems, Hamburg University of Technology, Eissendorfer Str. 40, 21073 Hamburg, Germany
 {\tt\small hamideh.saadabadi@tuhh.de}}%
\thanks{Gerald Gebhardt is with the Institute of Control Systems, Hamburg University of Technology, Eissendorfer Str. 40, 21073 Hamburg, Germany
 {\tt\small gerald.gebhardt@tuhh.de}}%
}

\newcommand\norm[1]{\left\lVert#1\right\rVert}

\ifthenelse{\equal{\TU}{true}}{
\headheight = 12pt
\headsep = 12pt
\topmargin = - 46pt
\fancyhf{}
\pagestyle{fancy}
\fancyhead[C]{This is an author’s copy supplementing a paper submitted to CDC 2022 and copyrighted by IEEE.}
}{
\pagestyle{empty}
}

\begin{document}

\maketitle

\ifthenelse{\equal{\TU}{true}}{
\thispagestyle{fancy}
\fancyhead[C]{This is an author’s copy supplementing a paper submitted to CDC 2022 and copyrighted by IEEE.}}{}

%
%
\begin{abstract}

This paper proposes a distributed scheme with different estimators for the event-triggered  formation control of polytopic homogeneously scheduled linear parameter-varying (LPV) multi-agent systems (MAS). Each agent consists of a time-triggered inner feedback loop and a larger event-triggered outer feedback loop to track a formation reference signal and reject input and output noise. If a local event-trigger condition is violated, the event-triggered outer feedback loop is closed through the communication network. The event-trigger condition is only based on locally available information. To design the controller, a synthesis problem is formulated as a linear matrix inequality of the size of a single agent under the assumption, that local estimators trigger intercommunication events with neighboring agents if the event-trigger condition is violated. The design procedure guarantees stability and bounded ${\ell_2}$-performance. Furthermore, the estimators are interchangeable for a given controller. We compare in simulation zero-order hold, open-loop estimation, and closed-loop estimation strategies. Simulation trials are carried out with non-holonomic dynamic unicycles modeled as polytopic LPV systems.

\end{abstract}

\section{Introduction}

Cooperative control of MAS has been widely studied, and the many applications include sensor networks, formation control of vehicles and
swarm robotics   \cite{c1}-\cite{c2}. 
Autonomous vehicles 
or 
mobile robots 
are often subject to non-holonomic constraints. 
The dynamics of such vehicles cannot be represented by linear time-invariant systems,  but they can be modeled as polytopic LPV systems~\cite{c3}.

The number of intercommunication events for MAS can be unnecessarily high if the agents communicate at  fixed time intervals determined by the sampling rate of a digital controller \cite{c4}-\cite{c5}. 
However, it can be reasonably reduced through the application of distributed event-triggered control strategies. In \cite{c6}-\cite{c23} the authors propose centralized and decentralized event-triggered formation control or consensus for LTI MAS. 
In \cite{Massioni} a MAS control problem is modelled via decomposable systems.

The event-triggered control of LPV systems is the subject of \cite{c24}-\cite{c39}. 
In \cite{c26}-\cite{c28} the design of an output feedback and event-triggered state feedback controller for discrete-time LPV systems with LMI conditions for bounded $\ell_2$-performance is studied. 
In \cite{c29} event-triggered $H_{\infty}$ state-feedback control for LPV systems with time triggered sensors is considered. 
This approach is extended in \cite{c30} to event-triggered output feedback control. 
The event-triggered control of continuous-time switched single loop LPV systems is considered in \cite{c31}-\cite{c32}. 
Event-triggered fault detection schemes are proposed in \cite{c33}-\cite{c34} for LPV systems.

In addition to event-triggered control, local state estimation can reduce the number of intercommunication events rapidly and reasonably \cite{c40}-\cite{c45}. 
In \cite{c40} sampled data is used for  event-triggered state estimation with bounded estimation error for a complex network. 
In \cite{c41} event-triggered, observer-based estimation is applied to reject disturbances and track references for discrete-time LTI systems. 
In \cite{c42} event-triggered distributed state estimation is used for a class of uncertain stochastic systems, which are subject to state-dependent noises and uncertainties. 
In \cite{c43} the optimal trade-off between the expected number of transmissions and the mean square estimation error is found over a finite horizon for an event-triggered estimation strategy for LTI systems. 
In \cite{c44} moving horizon event-triggered state estimation is used for an LTI MAS subject to noise and disturbances. 
In \cite{c45} an observer-based optimal event-triggered control strategy for LTI systems is proposed. 
In \cite{c46} event-triggered open-loop estimation is used for a undirected continuous LTI MAS.

To the best of the authors’ knowledge, so far no results on distributed event-triggered open-loop and fully connected closed-loop estimator-based control of LPV agent networks have been reported.

The contribution of this paper is a method for co-designing an event-triggered distributed LPV formation controller together with a trigger condition that depends only on locally available information, with guaranteed stability, $\ell_2$ performance, and interchangeable estimators. 
The gain-scheduled state feedback matrices are found via solving LMIs of the size of a single agent. 
The information transmitted to the neighbors depends on the applied estimator. 
The estimators require a certain communication structure and signal availability. 
Zero-order hold and open-loop estimation are considered for arbitrary communication graphs which are assumed to be undirected and connected, and closed-loop estimation is considered for a fully connected graph, in contrast to
\cite{c36}, where only the  zero-order hold estimation scheme is investigated.
The open-loop and closed-loop estimators also estimate the scheduling parameters.
For the proposed design procedure, it is assumed that the scheduling is homogeneous. 
This assumption is approximately satisfied for
non-holonomic agents which move in formation. 
A simulation example suggests that the proposed scheme will still work well even when this assumption is violated through a locally acting disturbance.


The remainder of this paper is organized as follows: The preliminaries and notation are given in section \ref{b:2}. 
The problem is formulated, and the estimators are proposed in section \ref{b:3}. 
Section \ref{b:4} presents a sufficient LMI condition for the considered problem. 
A non-holonomic system is introduced in 
section \ref{b:5} and the corresponding simulation results are given in section \ref{b:6}. 
Finally in section \ref{b:7} conclusions are presented.

\section{Preliminaries and Notation}\label{b:2}

For a signal $x_k$ taking values in  $\mathbb{R}^{n_x}$, 
the Euclidean 2-norm at time $k$ is $\norm{
x_k
}_{2} = \sqrt{x_k^\top x_k}
$
, and the signal ${\ell_2}$-norm is defined via $
\norm{
x_k
}_{\ell_2} = \sqrt{
{\sum^{\infty}_{k=0}
\norm{
x_k
}_{2}^2}.
}$ 
An $m \times n$ matrix, which consists only of zeros, is denoted by $0_{m \times n}$.
If the dimensions of a zero matrix are clear from context,
the index is omitted. The $i^{\text{th}}$ unit vector
$q_i\in\mathbb{R}^N$ is a column vector, which consists of zeros, except for the $i^{\text{th}}$ entry, which is equal to $1$. A column vector with $N$ rows, which entries are all equal to $1$, is denoted by $\mathbf{1}_N$. For two matrices 
$
A \in\mathbb{R}^{m \times n} :=[a_{ij}]
$ and $
B \in\mathbb{R}^{p \times q}
$
the Kronecker product between $A$ and $B$ is defined as $
A \otimes B :=[a_{ij}B]\in~\mathbb{R}^{mp \times nq}.
$
Properties of the Kronecker product are
$
(A \otimes B)^\top = A^\top \otimes B^\top 
$ and
$
(A \otimes B)(C \otimes D) = (AC) \otimes (BD).
$
A Kronecker product between a matrix $M\in \mathbb{R}^{N\times N}$ and the $n^\text{th}$ degree identity matrix $I_n$ is expressed as
$
M_{(n)} = M \otimes I_n.
$
Block diagonal matrices 
can be represented as $D = \text{diag}(D_1,\ldots,D_N)$, where $D_i$ is a scalar or a matrix.

The positive semi-definite and symmetric Laplacian $\mathcal{L}$ corresponds to the undirected and connected graph $\mathcal{G}$. The column space $\mathcal{R}(\mathbf{1}_N)$, which is the null space of the Laplacian $\mathcal{L}$, is the agreement space of the MAS and corresponds to the Laplacian's smallest and only non-positive eigenvalue $0=\lambda_1(\mathcal{L}) <\lambda_2(\mathcal{L}) \leq \ldots \leq \lambda_N(\mathcal{L})$ \cite{X}.

\section{Problem Formulation}\label{b:3}
\subsection{Agent Model}
Consider a group of $N$ agents with identical nonlinear dynamics, each modeled as polytopic LPV system $P(\theta_k^i)$ with state space realization
\begin{align}\label{eq:agent_model}
 x_{k+1}^i & = A(\theta_k^i)x_k^i + B_w(\theta_k^i)w_k^i + B_u(\theta_k^i)u_k^i, \notag\\
y_k^i & = C_y(\theta_k^i)x_k^i, 
\;
i=1,\ldots,N,
\end{align}
where $x^i_k\in \mathbb{R}^{n_x}$, $w^i_k \in \mathbb{R}^{n_w}$, $u^i_k \in \mathbb{R}^{n_u}$ and $y^i_k \in \mathbb{R}^{n_y}$, denote the state, perturbation, input and transmitted output at time $k$. 
The system $\check{P}(\theta_k)$ containing the dynamics of all agents is expressed via
\begin{align}
x_{k+1}&=\check{A}(\theta_{k})x_k+\check{B}_u(\theta_{k})u_k+\check{B}_w(\theta_{k})w_k,\notag
\\
y_k&=\check{C}_y(\theta_{k})x_k,\label{e:SS}
\end{align}
where stacked vectors, e.g. $
x_k = \left[ (x_k^1)^\top (x_k^2)^\top \cdots (x_k^N)^\top \right]^\top,
$
and diagonal matrices, e.g. 
\[
\check{A}(\theta_{k}) = \text{diag}\left(A(\theta^1_{k}),A(\theta^2_{k}), \ldots, A(\theta^N_{k})\right),
\]
are used.
The model matrices, e.g. $A(\theta^{i}_{k})$ and $B_w(\theta^{i}_{k})$, depend affinely on the time-varying vector of scheduling variables $\theta^{i}_{k}$ of agent $i$. 
The scheduling variables $\theta^{i}_{k}$ are restricted to a compact set $\Theta$ via $\theta^{i}_{k} \in \Theta \subset \mathbb{R}^{n_\theta}$ at all times. 
We assume that the parameter set 
\[
\Theta = \left\{ \theta \in \mathbb{R}^{n_\theta} | \theta = \sum_{l=1}^s \alpha_l \theta_{l}, \sum_{l=1}^s \alpha_l= 1, \alpha_l \geq 0\right\}
\]
is represented as a polytope in terms of vertex vectors $\theta_{l}~\in~\mathbb{R}^{n_\theta}$. This restriction is represented via $\theta \in
\mathcal{F}_{\Theta}$. 
Defining vertex model matrices, e.g. $A^l=A(\theta_{l})$, the LPV model matrices for agent $i$ can be expressed in terms of the convex coordinates $\alpha_l(\theta_k^i)$ as 
 \begin{equation}\label{eq:agent_polytope} 
\left[ \begin{array}{ccc}
A(\theta^i_{k}) & B_w(\theta^i_{k}) & B_u(\theta^i_{k})\\
C_y(\theta^i_{k}) &0&0\\ 
\end{array} 
\right] = \sum_{l=1}^{s} \alpha_l(\theta^i_{k}) \left[ \begin{array}{ccc}
A^l & B_w^l&B_u^l \\
C_y^l& 0&0
\end{array} \right].
\end{equation}
\subsection{Event-Triggered Formation Control}
\begin{figure}[ht]
\centering
\def\svgwidth{\columnwidth}
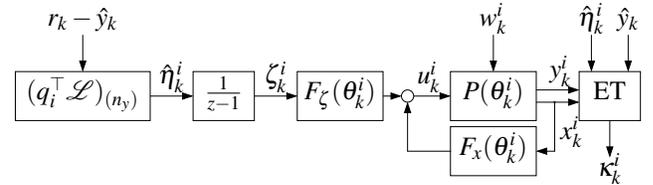
\caption{Control structure for a single agent}
\label{f:singleagent}
\end{figure}

\begin{figure}[ht]
\centering
\def\svgwidth{\columnwidth}
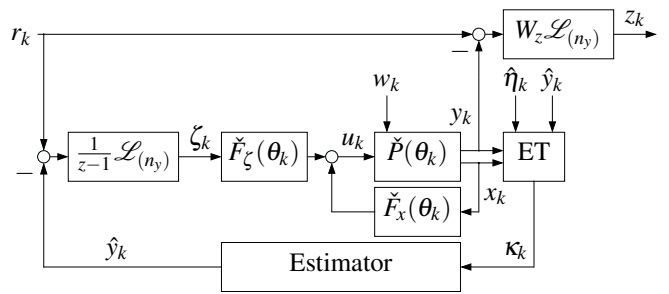
\caption{Control structure of the closed-loop network with performance output and estimator}
\label{f:network}
\end{figure}

The group of $N$ agents, governed by~(\ref{e:SS}), can communicate through an undirected and connected graph~$\mathcal{G}$. 
Every agent estimates the output $\hat{y}^j_k$ of its neighbors $j \in \mathcal{N}_i$ and its own output $\hat{y}^i_k$ in the same way as its neighbors do, to calculate the estimated formation error
\begin{equation}
\hat{\eta}^i_k = 
\sum_{j \in {\cal N}_i} ( (r^i_k - \hat{y}_k^i)-(r^j_k-\hat{y}_k^j) )=
(q_i^\top\mathcal{L})_{(n_y)}(r_k - \hat{y}_k)\label{e:eta}
\end{equation}
locally,
where $r_k^i$ is the reference, which should be tracked.
The control structure for agent~$i$ is shown in Figure~\ref{f:singleagent}.
 The estimated formation error $\hat{\eta}_k$ is used to approximate the formation error 
$
{\eta}_k = 
\mathcal{L}_{(n_y)}(r_k - {y}_k).
$
Figure \ref{f:network} shows the closed-loop network, where the cost $z_k$ and also estimators are pictured. 

%
The "ET"-block in Figure \ref{f:singleagent} evaluates the local trigger condition 
\begin{equation}
(e^i_{k})^\top e^i_{k} \leq \sigma ({\hat{\eta}_{k}^i})^\top \hat{\eta}_{k}^i, \label{e:ETC}
\end{equation} 
where 
$
e^i_k = \hat{y}^i_k - y^i_k $
is the estimation error and $\sigma > 0$ is the trigger level. 
If the current estimate $\hat{y}_k^i$ violates the event-trigger condition, a message $\kappa^i_k$ is sent. This sets the estimation error $e_k^i$ to zero. A message consists of necessary information to locally formulate new identical estimates $\hat{y}^i_{k+1}$, $\hat{y}^i_{k+2}$, $\ldots$ by every agent $j \in \mathcal{N}_i \cup i$. 
 Three different estimators are presented in subsection \ref{b:esti}.
 The integrated formation error 
\begin{equation}
\zeta^i_{k+1} = \zeta^i_k + \hat{\eta}^i_k \label{e:zeta}
\end{equation}
is used together with the agent's states $x^i_k$ for gain-scheduled state feedback via 
\begin{equation}
 u_k^i = F_{\zeta}(\theta_k^i) {\zeta}_k^i + {F}_x(\theta_k^i) {x}_k^i.\label{e:u}
\end{equation}
Since the agents are affine in $\theta_k^i$, affine parameter dependence is imposed on the controllers, too, i.e. 
\begin{equation}
F_x(\theta_k^i) = \sum_{l=1}^s \alpha_l(\theta_k^i)F_x^l, \; F_\zeta(\theta_k^i) = \sum_{l=1}^s \alpha_l(\theta_k^i)F_\zeta^l, 
\label{e:F}
\end{equation}
where $F_x^l=F_x(\theta_{l})$ and $F_\zeta^l=F_\zeta(\theta_{l})$ denote the vertex controllers at vertex $l$. Thus, the input signal can be expressed via
\begin{equation}
 u_k = \check{F}_{\zeta}(\theta_k) {\zeta}_k + \check{F}_x(\theta_k) {x}_k.\label{e:U}
\end{equation}
Since the control objective is to design a controller which follows a reference input~$r_k$ and rejects the disturbance~$w_k$ the control performance can be measured with the performance output
\begin{align}
z_k&= w_{z}(k) \ast \mathcal{L}_{(n_y)}(r_k-y_k)\label{e:performance},
\end{align}
where $(\ast)$ denotes the discrete convolution and $W_z(z)=\frac{\beta_z}{z+\alpha_z}$ is the $z$-transformed filter $w_z(k)$ with variables~$\alpha_z$ and~$\beta_z$ as design parameters. The parameter~$\alpha_z$ is used to set a weight on the latest cost function value. 
With~(\ref{e:SS}) the performance output~$z_k$ can be expressed as state space model
\begin{align}
\tilde{z}_{k+1} &= - \alpha_z\tilde{z}_k + \beta_zr_k - \beta_z\check{C}_y( \theta_k)x_k ,\;
z_k=\mathcal{L}\tilde{z}_{k}, \label{e:tildez}
\end{align}
with the local performance~$\tilde{z}_{k}$ as state vector.~(\ref{eq:agent_model}),~(\ref{e:eta}) and~(\ref{e:zeta})
 lead to
\begin{align}
\hat{\eta}_k & = \mathcal{L}_{(n_y)}(r_k-\check{C}_y( \theta_k)x_k-e_k),\notag\\
\zeta_{k+1} & = \zeta_k + \mathcal{L}_{(n_y)}(r_k-\check{C}_y( \theta_k)x_k-e_k).\label{e:zeta2}
\end{align}
Thus,~(\ref{eq:agent_model}),~(\ref{e:U}),~(\ref{e:tildez})
 and~(\ref{e:zeta2}) can be combined to form 
\begin{align}
&\left[
\begin{array}{ccc|c|c}
x_{k+1}^\top
&
\zeta_{k+1}^\top
&
\tilde{z}_{k+1}^\top
&
\hat{\eta}_k^\top
&
z_k^\top
\end{array}
\right]^\top =
\notag
\\
&
\left[
\begin{array}{ccc|c|ccc}
\check{A}+\check{B}_u\check{F}_x &\check{B}_u\check{F}_{\zeta} &0 & 0 & 0 &\check{B}_w 
\\
-\mathcal{L}_{(n_y)}\check{C}_y & I_{Nn_y} & 0 &-\mathcal{L}_{(n_y)} &\mathcal{L}_{(n_y)} & 0
\\
-\beta_z\check{C}_y & 0 &-\alpha_z I_{Nn_y} & 0 &\beta_z I_{Nn_y} & 0
\\ \hline
-\mathcal{L}_{(n_y)}\check{C}_y & 0 & 0 &-\mathcal{L}_{(n_y)} &\mathcal{L}_{(n_y)} & 0
\\ \hline
0 &0 &\mathcal{L}_{(n_y)} &0 &0 &0
\end{array}
\right]\notag
\\ \cdot & \left[
\begin{array}{ccc|c|cc}
x_{k}^\top &
\zeta_{k}^\top &
\tilde{z}_{k}^\top &
e_k^\top &
r_k^\top &
w_k^\top
\end{array}
\right]^\top,
\label{e:Dynbig}
\end{align} 
where the dependence on the scheduling parameter $\theta_k$ is omitted for ease of notation.~(\ref{e:Dynbig}) can be abbreviated via
\begin{align}
\underbrace{
\left[
\begin{array}{c}
\psi_{k+1} \\
\hline
\hat{\eta}_k \\
\hline
z_k
\end{array}
\right]}_{=\nu_k}
=
\underbrace{ 
\left[
\begin{array}{c|c|c}
\bar{A}_{\text{CL}} & \bar{B}_e & \bar{B}_{{{f}}} \\
\hline
\bar{C}_{\hat{\eta}} & \bar{D}_{\hat{\eta}e} & \bar{D}_{\hat{\eta}{{f}}} \\
\hline
\bar{C}_{z} & \bar{D}_{ze} & \bar{D}_{z{{f}}}
\end{array}
\right]}_{=\bar{H}( \theta_k)}
\underbrace{
\left[
\begin{array}{c}
\psi_{k} \\
\hline
e_k \\
\hline
f_k
\end{array}
\right]}_{=\phi_k},
\label{e:Dynsmll}
\end{align}
where
\[
\psi_k=\left[x_k^\top 
\zeta_k^\top 
\tilde{z}_k^\top\right]^\top\in\mathbb{R}^{N\cdot n_{\psi}},\;
f_k =\left [r_k^\top w_k^\top \right]^\top\in\mathbb{R}^{N\cdot n_f},
\]
$n_{\psi}=n_x+n_y+n_z$ and $n_f = n_y+n_w$. 
The horizontal~$(-)$ and vertical~$(\, |\,)$ lines make the correlations between~(\ref{e:Dynbig}) and~(\ref{e:Dynsmll}) unambiguous. The closed-loop matrix 
\begin{align}
\bar{A}_{\text{CL}}&=\underbrace{\left[
\begin{array}{ccc}
\check{A} & 0 & 0 \\
-\mathcal{L}_{(n_y)}\check{C}_y & I_{Nn_y} & 0 \\
-\beta_z\check{C}_y & 0 &-\alpha_z I_{Nn_y}
\end{array}
\right]
}_{=\bar{A}_{\text{OL}}}
+
\underbrace{\left[
\begin{array}{c}
\check{B}_u \\
0 \\
0
\end{array}
\right]}_{=\bar{B}_F}
\underbrace{\left[
\begin{array}{c}
\check{F}_x^\top \\
\check{F}_{\zeta}^\top \\
0
\end{array}
\right]^\top}_{=\bar{F}}
\label{e:INny}
\end{align}
introduces feedback via $\bar{F}$. In equation~(\ref{e:INny}), $I_{Nn_y}$ denotes the identity matrix of degree $N\cdot n_y$.

\subsection{Estimation}\label{b:esti}
The proposed zero-order hold estimation, open-loop estimation and closed-loop estimation schemes require different information to formulate new estimates locally.
\subsubsection{Zero-Order Hold Estimation}
For zero-order hold estimation, the necessary information is the output $\kappa^i_k=y^i_k$, since $\hat{y}^i_{k+1} = \hat{y}^i_{k}$ is the estimate, if no event-triggering occurs at agent $i$.
This estimate is updated to $\hat{y}^i_{k}=y^i_k=\kappa^i_k$, if an event is triggered at time $k$.
\subsubsection{Open-Loop Estimation}
For open-loop estimation, agent $i$ sends a message $\kappa^i_k=[(x_k^i)^\top(\theta^i_k)^\top]^\top$ to its neighbors when the event-trigger condition is violated, since the local estimation is based upon the prediction of the next scheduling parameter $\hat{\theta}^i_{k+1} = f(\hat{\theta}^i_{k}, \hat{x}^i_{k})$ and the output prediction via
\begin{align}
\left[
\begin{array}{c}
\hat{x}_{k+1}
\\
\hline
\hat{y}_k
\end{array}
\right]
= 
\left[
\begin{array}{c}
A( \hat{\theta}^i_k)
\\
\hline
C_y(\hat{\theta}^i_k)
\end{array}
\right]
\hat{x}^i_k.
\end{align}
The function $f(\hat{\theta}^i_{k}, \hat{x}^i_{k})$ is a combination of the locally available parameters $\hat{\theta}^i_{k}$ and $\hat{x}^i_{k}$ to predict the next scheduling parameter $ \hat{\theta}^i_{k+1}$.

\subsubsection{Closed-Loop Estimation for Fully Connected Network}
For the proposed closed-loop estimator, the network needs to be fully connected, i.e. 
$\mathcal{L} = NI_N-\mathbf{1}_N\mathbf{1}_N^\top$, since every agent estimates the full network and is required to produce the same estimate $\hat{y}_k$ as its neighbors.
Furthermore, the necessary information for this estimation scheme are 
$\hat{\theta}_{k},$ $\hat{x}_{k}$, $\hat{\zeta}_{k}$ and $r_k$,
since output estimates $\hat{y}_k$ are generated via $\hat{\theta}_{k+1} = f(\hat{\theta}_{k}, \hat{x}_{k},\hat{\zeta}_{k},r_k)$ and
\begin{align}
&
\left[
\begin{array}{cc|c}
\hat{x}_{k+1}^\top
&
\hat{\zeta}_{k+1}^\top
&
\hat{y}_k^\top
\end{array}
\right]^\top =
\notag
\\
&
\left[
\begin{array}{cc|ccccc}
\check{A}( \hat{\theta}_k)+\check{B}_u( \hat{\theta}_k)\check{F}_x( \hat{\theta}_k) &\check{B}_u( \hat{\theta}_k)\check{F}_{\zeta}( \hat{\theta}_k) &0 
\\
-\mathcal{L}_{(n_y)}\check{C}_y( \hat{\theta}_k) & I_{Nn_y}&\mathcal{L}_{(n_y)} 
\\\hline
\mathcal{L}_{(n_y)}\check{C}_y( \hat{\theta}_k) & 0 & 0 
\end{array}
\right]
\left[
\begin{array}{c}
\hat{x}_{k}
\\
\hat{\zeta}_{k}
\\
\hline
r_k
\end{array}
\right]
.
\label{e:CLE}
\end{align} 
Thus, if an event is triggered at agent $i$ and time $k$, agent $i$ sends a message $\kappa^i_k=[({\theta}^i_{k})^\top ({x}^i_{k})^\top ({\zeta}^i_{k})^\top]^\top$ to the other agents.
\ifthenelse{\equal{\short}{true}}{}{
Table \ref{t:EstiCons} summarizes the constraints regarding the communication graph and signal availability for the proposed estimators.

\begin{table}[ht]
	\caption{Estimator Constraints} 	\label{t:EstiCons}
	\begin{center}
	\begin{tabular}{|r|ll|}
	\hline
	Estimator & $\mathcal{G}$ & $\kappa^i_k$
	\\
	\hline
	ZOH & undirected, connected & $y^i_k$
	\\
	OLE & undirected, connected & ${\theta}^i_{k}, {x}^i_{k}$
	\\
	CLE & fully connected & ${\theta}^i_{k}, {x}^i_{k}, {\zeta}^i_{k}$
	\\
	\hline
	\end{tabular}
	\end{center}
\end{table}
}

\subsection{Problem Formulation}
The problem addressed in this paper is the following:
For a given positive constant $\gamma$, a network with dynamics
governed by~(\ref{e:Dynbig}) and~(\ref{e:Dynsmll}) 
and a given estimator, which ensures that~(\ref{e:ETC}) is true, 
find the scheduled state feedback gain matrices $F_x(\theta_k^i)$ and $F_\zeta(\theta_k^i)$ in (\ref{e:u}), such that the group of $N$ agents is stable and satisfies
\begin{equation}
\left.
\norm{ 
T_{z{{f}}} 
}
_{\ell_2}
\right\vert_{x_0=0}
= 
\sup_{\theta \in
\mathcal{F}_{\Theta}}
\sup_{{{f}}\neq 0}
\left.
\frac{
\norm{
z_k
}
_{\ell_2}}{
\norm{
{{f}}_k
}
_{\ell_2}}
\right\vert_{x_0=0}
\leq \gamma. \label{e:T1}
\end{equation}
\section{Controller Synthesis}\label{b:4}
{\em Assumption 1:}
The scheduling in the group is homogeneous, i.e. $\theta_k^i = \theta_k^j, \; \forall \; 1 \leq i\leq N, 1 \leq j \leq N \text{ and } k \geq 1.$

The matrix inequality
\begin{eqnarray}
\left[
\begin{array}{cccccc}
 {G^l}^\top+{G^l}-{S} \\
0 & I_{n_y}& & & \ast \\
0 & 0 &t I_{n_f}\\
{A}^{li}_{\text{OL}}{G}^l+B_F^lK^l & {B}^{i}_e & {B}^{li}_{{{f}}} 
& {S}
\\
{C}^{li}_{\hat{\eta}}{G}^l & {D}^{i}_{\hat{\eta}e} & {D}^{i}_{\hat{\eta}{{f}}} &
0 & \sigma_x I_{n_y}
\\
{C}^{i}_{z}{G}^l & {D}_{ze} & {D}_{z{{f}}} &
0 & 0 & I_{n_y}
\end{array}
\right] > 0, \label{e:M4i1}
\end{eqnarray}
where 
\begin{align}
&G^l = \text{diag}(G^l_1,G^l_2),\:
G_1^l = \left[
\begin{array}{cc}
{G}_{x}^l & {G}_{x \zeta}^l 
\\
{G}_{\zeta x}^l & {G}_{\zeta}^l
\end{array}
\right],
\\
& K^l = [K^l_1 0_{n_u \times n_y}]
,
\:
K_1^l = [
{K}_{x}^l 
\:
{K}_{x \zeta}^l
],\:
B_F^l = \label{e:M4i2}
\left[
\begin{array}{c}
B_u^l \\
0_{2\cdot n_y \times n_u} \\
\end{array}
\right],
\\
& n_f = n_y + n_w,\:
t = \gamma^2,\:
\sigma_x = \sigma^{-1},
\\
&
\left[
\begin{array}{c|c|c}
{A}^{li}_{\text{OL}} & {B}^{i}_e & {B}^{li}_{{{f}}} \\
\hline
{C}^{li}_{\hat{\eta}} & {D}^{i}_{\hat{\eta}e} & {D}^{i}_{\hat{\eta}{{f}}} \\
\hline
{C}^{i}_{z} & {D}_{ze} & {D}_{z{{f}}}
\end{array}
\right] = \notag
\\
&
\left[
\begin{array}{ccc|c|ccc}
{A^l} &0 &0 & 0 & 0 & {B}_w^l \\
-\lambda_i{C}^l_y & I_{n_y} & 0 &-\lambda_i I_{n_y} &\lambda_i I_{n_y} & 0\\
-\beta_z{C}^l_y & 0 &-\alpha_z I_{n_y} & 0 &\beta_z I_{n_y} & 0\\
\hline
-\lambda_i{C}^l_y & 0 & 0 &-\lambda_i I_{n_y} &\lambda_i I_{n_y} & 0\\
\hline
0 &0 &\lambda_i I_{n_y} &0 &0 &0 
\end{array}
\right] \label{e:M4i3}
\end{align}
is linear in $S$, $G^l$, $K^l$, $t$ and $\sigma_x$. In~(\ref{e:M4i1}), ($\ast$) denotes a matrix block that can be inferred by symmetry.

\begin{thm}
The closed-loop MAS defined by~(\ref{e:Dynbig}) and~(\ref{e:Dynsmll}),
with an estimator and event-triggering mechanism, which ensures that~(\ref{e:ETC}) is true and distributed feedback governed by~(\ref{e:F}), where 
$(F^l_x F^l_\zeta) = K^l_1{G^l_1}^{-1}$
is stable and satisfies (\ref{e:T1}), 
if there exist a symmetric positive-definite matrix
$S\in\mathbb{R}^{n_{\psi} \times n_{\psi}}$ and matrices 
$K^l_1\in\mathbb{R}^{n_{u} \times (n_y+n_x)}$,
$G^l_1\in\mathbb{R}^{(n_y+n_x)\times (n_y+n_x)}$ and 
$G^l_2\in\mathbb{R}^{n_y\times n_y}$ that satisfy (\ref{e:M4i1}) for $l=1,2,\ldots,s$ and $i = 2,N$. 
\end{thm}

{\em Proof:} \ifthenelse{\equal{\TU}{true}}{See appendix}{
Due too spacial limitations the corresponding proof is omitted, but can be attained via.
}

Note that the assumption of homogeneous scheduling may be violated, especially when a perturbation acts on an individual agent. The simulations shown below suggest, that even then the proposed scheme performs well.

\section{LPV Representation of a Non-Holonomic Vehicle}\label{b:5}
To motivate the use of LPV models as agents in formation control problems, unicycles are used as an example for non-holonomic systems.
The position of the agent in a plane is given by the Cartesian coordinates $x(t)$ and $y(t)$ and its orientation by the angle $\phi(t)$.
The non-holonomic constraint on the mobile robot shown in Figure \ref{f:rolling} is that only acceleration via a force $f$ and steering via a torque $\tau$ is allowed.
\subsection{Non-holonomic system with handle point}
The unicycle's dynamic equations are 
\[
\dot{x}=v \cos(\phi), \; \dot{y}=v \sin(\phi), \; \dot{v} = \frac{f}{m},\; \dot{\phi} = \omega, \dot{\omega} = \frac{\tau}{I},
\]
where $m$ is the unicycle's mass and $I$ is its moment of inertia.
\begin{figure}[ht]
\centering
\def\svgwidth{\columnwidth}
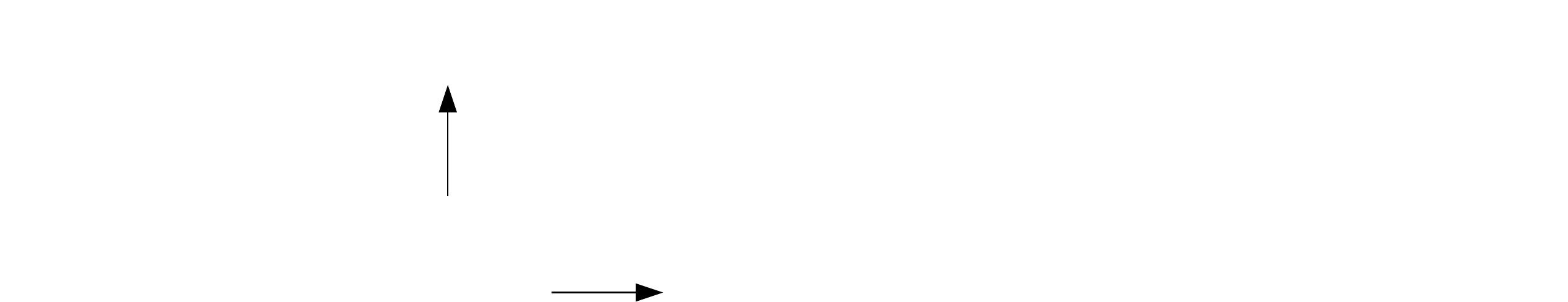
\caption{Dynamic unicycle with handle}
\label{f:rolling}
\end{figure}
Figure \ref{f:rolling} shows the unicycle with a fictitious handle point at position ($x^d,y^d$), which is a distance $d$ apart in $\tilde{x}$-direction from the unicycles center of mass at position ($x,y$). The handle point is introduced to improve the system's controllability \cite{Y}. The transformation matrix 
\begin{equation}
T_\phi = \left[ \begin{array}{cc}
\cos \phi & \sin \phi\\
-\sin \phi & \cos \phi\\
\end{array} \right]
\end{equation}
is used to transform from $xy$-coordinates to $\tilde{x}\tilde{y}$-coordinates.
An LPV model of the handle is then obtained by writing the state space model as 
\begin{align}
\notag
&\left[
(\dot{\tilde{x}}^d)^\top
(\dot{\tilde{y}}^d)^\top
\dot{v}_n^\top
\dot{v}_t^\top
\right]^\top=
\\
&
\left[ \begin{array}{cccc}
0 & \frac{v_t}{d} & 1& 0 \\
-\frac{v_t}{d} & 0 & 0 & 1 \\
0 & 0 & 0 & 0 \\
0 & 0 & 0 & 0 
\end{array} \right] 
\left[ \begin{array}{c}
\tilde{x}^d \\ \tilde{y}^d \\ v_n \\ v_t
\end{array} \right]
 + 
 \left[ \begin{array}{cc}
0 & 0 \\
0 & 0 \\
\frac{1}{m} & 0 \\
0 & \frac{d}{I} 
\end{array} \right]
\left[ \begin{array}{c}
f \\ \tau
\end{array} \right], \label{e:LPV}
\end{align}
where $v_n$ and $v_t$ are the normal and lateral velocities and the only scheduling parameter is $v_t = \theta$. Note, that~(\ref{e:LPV}) has the form
\begin{equation}\label{eq:5}
\dot{x}(t) = A(\theta(t)) x(t) + B_uu(t)
\end{equation}
and 
that $A(\theta(t))$ is affine in $\theta$. Furthermore, the matrices $B_w$, $B_u$ and $C_y$ are independent of the scheduling parameter. Thus, the design procedure proposed in section \ref{b:4} is applicable.
Euler's discretization is used to discretize the model with sampling time $T_{s}$ via
 \begin{equation}
{x}_{k+1} =(I+ T_s\cdot A(\theta_k) ) x_k + (T_s\cdot B_u) u_k
\end{equation}
for the simulations.
\section{Simulation Results}\label{b:6}
\begin{figure}[ht]
		\centering
	 	\input{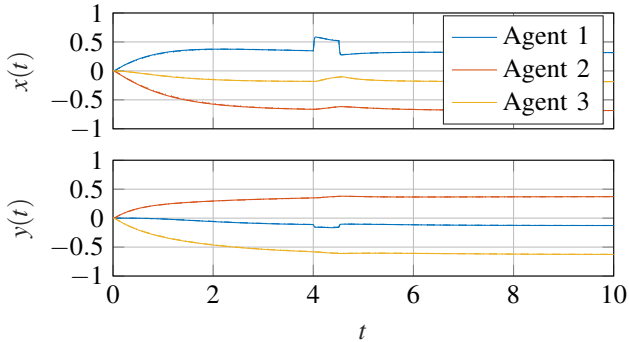} 
	 	\caption{Spatial response for ZOH, OLE and CLE}
	 	\label{f:XY} 
\end{figure}

\begin{figure}[ht]
		\centering
	 	\input{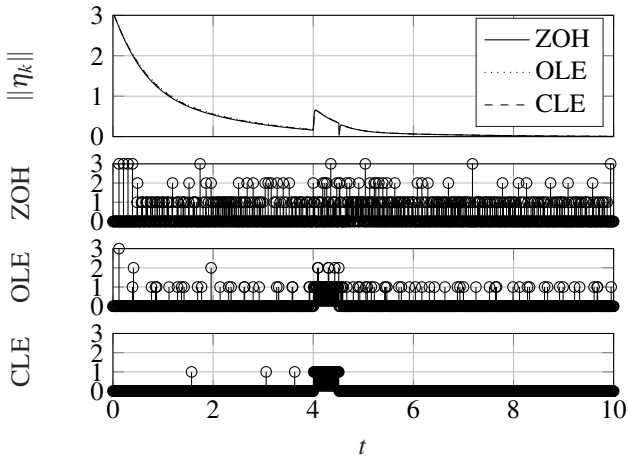} 
	 	\caption{Formation error and simultaneous trigger events for ZOH, OLE and CLE}
	 	\label{f:EtaTrig} 
\end{figure}

To show the effectiveness of the proposed method, a simulation example is provided \cite{Z}. For the simulations a network of three agents with complete communication graph and corresponding Laplacian $\mathcal{L} = 3I_3-\mathbf{1}_3\mathbf{1}_3^\top$ is used. The reference is 
$
r_k = [1,0,0,0.5,0.5,-0.5]^\top,
$
the trigger level ist 
$\sigma = 10^{-3}$
and the sampling time is $T_s = 0.01$.
Between the times $t=4$ and $t=4.5$ an $\tilde{x}$-directional input disturbance $w_k^1$ is applied to the agent 1. The simulation results are provided by Figures \ref{f:XY} and \ref{f:EtaTrig}.
Figure \ref{f:XY} shows the $xy$-directional time response of the three agents for zero-order hold, open-loop and closed loop estimation. The spatial responses are overlapping each other in Figure \ref{f:XY} for the three estimators.
Figure \ref{f:EtaTrig} displays the Euclidean 2-norm of the formation error~$\eta_k$ and the amount of simultaneous trigger instants per time instant. The differences in the spatial response are negligible, but the amount of trigger events varies significantly. This result is also supported by Table \ref{t:EstiPerf}.
\begin{table}[ht]
	\caption{Estimator Performance} 	\label{t:EstiPerf}
	\begin{center}
	\begin{tabular}{|r|lll|}
	\hline
	Estimator & $\overline{\norm{\eta}_2}$ & Trigger events & Trigger rate
	\\
	\hline
	ZOH & 0.3852 & 362 & 0.1205
	\\
	OLE & 0.3995 & 131 & 0.0436
	\\
	CLE & 0.3923 & 54 & 0.0180
	\\
	\hline
	\end{tabular}
	\end{center}
\end{table}
The mean of the formation error norm 
$
\overline{\norm{\eta}_2}
=\sum_{k=1}^{\frac{30}{T_s}} \frac{T_s\norm{\eta_k}_2}{30}
$
in Table \ref{t:EstiPerf} does not differ significantly for the three proposed estimators. However, Table \ref{t:EstiPerf} also shows the amount of trigger events and the trigger rate, which is the amount of actual trigger events divided by the amount of possible trigger events. The amount of trigger events and the trigger rate is almost half as large for OLE compared to ZOH and more than seven times smaller for CLE compared to ZOH in the considered simulation.

\section{Conclusions}\label{b:7}
This paper proposes a distributed event-triggered formation control scheme for polytopic LPV MAS, that guarantees stability and a level of control performance for the network and is suitable for different local estimation schemes. 
The synthesis problem is formulated as an LMI problem of the size of a single agent. 
The estimators reduce the communication load and are interchangeable for given control matrices. 
 A simulation example illustrates the practicality of the proposed method and compares the proposed estimators. 

%
%
%

%
%
%
%
%
%
%
%
\ifthenelse{\equal{\TU}{true}}{
{\em Proof of Theorem 1:}

Consider the Lyapunov function candidate
$$V(x_k,
\zeta_k,\tilde{z}_k)=\left[
\begin{array}{c}
x_k
\\
\zeta_k
\\
\tilde{z}_k
\end{array}
\right]^\top 
\underbrace{
\left[
\begin{array}{ccc}
\check{P}_{x} & \check{P}_{x \zeta} & \check{P}_{x \tilde{z}} 
\\
\check{P}_{\zeta x} & \check{P}_{\zeta} & \check{P}_{\zeta\tilde{z}} 
\\
\check{P}_{\tilde{z}x} & \check{P}_{\tilde{z}\zeta} & \check{P}_{\tilde{z}} 
\\
\end{array}
\right]
}_{=\bar{P}}
\left[
\begin{array}{c}
x_k
\\
\zeta_k
\\
\tilde{z}_k
\end{array}
\right],$$
where $\bar{P}=\bar{P}^\top>0$. Stability, $\ell_2$ performance due to~(\ref{e:T1}) and the event-trigger condition via~(\ref{e:ETC}) are combined to 
\begin{align}\label{eq:et_stab}
\ {{\psi}}^\top_{k+1} \bar{P} {\psi}_{k+1}
 -{{\psi}}^\top_k\bar{P}{\psi}_k 
 & \leq \notag
 \\
 -{{z}_k}^\top{{z}_k}+\gamma^2 {f_k}^\top {f_k}
 -\sigma {\hat{{\eta}}^\top_k}\hat{{\eta}}_k+{e}_{k}^\top{e}_{k} &.
\end{align}
Applying~(\ref{e:Dynsmll}) leads to
\begin{align}
0 &
\geq
\phi_k^\top 
\big((
\bar{H}^\top(\theta_k)
\bar{R}_2 \bar{H}(\theta_k) 
-
\bar{R}_1
\big)\phi_k 
, \notag
\\
\bar{R}_1 &= \text{diag}(
\bar{P},I_{Nn_y},\gamma^2 I_{Nn_f}
), \notag
\\
\bar{R}_2 &=
\text{diag}(
\bar{P},\sigma I_{Nn_y}, I_{Nn_y}
), 
\end{align}
where $n_f = n_y+n_w$. Using a Schur complement results in
\begin{equation}
\left[
\begin{array}{cc}
\bar{R}_1 & 
\bar{H}^\top(\theta_k)
\bar{R}_2^\top
 \\
\bar{R}_2 \bar{H}(\theta_k) & \bar{R}_2
\end{array}
\right] \geq 0,
\end{equation}
which is a non-linear matrix inequality. This inequality is linearized via multiplying it from the right by $\text{diag}(\bar{D}(\theta_{k}),
\bar{R}_2^{-1}),$
where $$\bar{D}(\theta_{k}) = \text{diag}(\bar{G}(\theta_{k}),I_{Nn_y+Nn_w})$$
and
$$\bar{G}(\theta_{k}) = \text{diag}\left(
\left[
\begin{array}{cc}
\check{G}_{x}(\theta_{k}) & \check{G}_{x \zeta}(\theta_{k}) 
\\
\check{G}_{\zeta x}(\theta_{k}) & \check{G}_{\zeta}(\theta_{k}) 
\end{array}
\right],
\check{G}_{{2}}(\theta_{k}\right) $$
 and from the left by its transpose, using the inequality
$
\bar{S}
\leq
\bar{G}(\theta_{k})+\bar{G}(\theta_{k})^\top- \bar{S},
$
where $\bar{S} = \bar{P}^{-1}$,
 to get the more conservative but linear matrix inequality
\begin{equation}
\left[
\begin{array}{cc}
\bar{M}_{11}(\theta_{k}) & \bar{M}_{21}^\top(\theta_{k}) \\
\bar{M}_{21}(\theta_{k}) & \bar{M}_{22}
\end{array}
\right] \geq 0, \label{e:M}
\end{equation} 
 where
\begin{align}
\bar{M}_{11}(\theta_{k})& = \text{diag}\big(
\bar{G}(\theta_{k})+\bar{G}(\theta_{k})^\top- \bar{S},I_{Nn_y},t I_{Nn_f}
\big), 
\notag
\\
\bar{M}_{21}(\theta_{k}) &= \left[
\begin{array}{ccc}
\bar{A}_{\text{OL}}\bar{G}+\bar{B}_F\bar{K} & \bar{B}_e & \bar{B}_{{{f}}} \\
\bar{C}_{\hat{\eta}}\bar{G} & \bar{D}_{\hat{\eta}e} & \bar{D}_{\hat{\eta}{{f}}} \\
\bar{C}_{z}\bar{G} & \bar{D}_{ze} & \bar{D}_{z{{f}}}
\end{array}
\right](\theta_{k}),
\notag
\\
\bar{M}_{22} &= \text{diag}(
\bar{S},\sigma_x I_{Nn_y}, I_{Nn_y}
),
\notag
\\
\bar{K}(\theta_{k}) &= [\check{K}_{x}(\theta_{k})\: \check{K}_{x \zeta}(\theta_{k})\: 0_{Nn_u \times Nn_y}], 
\:
t = \gamma^2,
\:\sigma_x = \sigma^{-1}.
\end{align} 
Under the assumption of homogeneous scheduling (Assumption 1), the matrix inequality can be diagonalized, since $$\theta_k = \mathbf{1}_N \otimes \theta_k^1$$ implies applications of the Kronecker product, e.g. 
$$\check{A}(\theta_k) = I_N \otimes A(\theta_k^1),$$ and the transformation 
$$Z_{(n_x)}^\top\mathcal{L}_{(n_x)}Z_{(n_x)} = \text{diag}(0,\lambda_2,\ldots,\lambda_N)\otimes I_{n_x}$$ has no effect on e.g. $\check{A}(\theta_k)$
 for orthonormal transformation matrices $Z$, i.e. $Z_{(n_x)}^\top \check{A}(\theta_k) Z_{(n_x)} = \check{A}(\theta_k)$, due to properties of the Kronecker product. Thus, the transformation matrix 
 \begin{align}
\bar{T}= \text{diag}
\big(
Z_{(n_{x})}, 
Z_{(n_{y})}, 
Z_{(n_{y})}, 
Z_{(n_{y})}, 
Z_{(n_{y})}, 
Z_{(n_{w})}, 
\notag
\\
Z_{(n_{x})}, 
Z_{(n_{y})}, 
Z_{(n_{y})}, 
Z_{(n_{y})}, 
Z_{(n_{y})} 
\big)
 \end{align}
 brings~(\ref{e:M}) into diagonal form, where entries are dependent on $\lambda_i$ and ordered from $i=1$ to $i=N$, displayed for the vertex model matrices by~(\ref{e:M4i1}).
The inequality~(\ref{e:M4i1}) needs to be feasible for $l=1,\ldots,s$ for $\lambda_2$ and $\lambda_N$, since $\theta^{i}_{k} \in \Theta$ and
the matrices are affine in $\lambda_i$ and $\lambda_1$ corresponds to the agreement space.}{}

\end{document}